\documentstyle[emulateapj]{article}
\def\keywords{}
\def\acknowledgements{}

\begin{document}

\def\nth{n_{\rm th}}
\def\nobs{n_{\rm obs}}
\def\dmin{d_{\rm min}}
\def\macho{{\sc macho}}
\def\newpage{\vfill\eject}
\def\vs{\vskip 0.2truein}
\def\gnu{\Gamma_\nu}
\def\fnu {{\cal F_\nu}}
\def\mass{m}
\def\lum{{\cal L}}
\def\imf{\xi(\mass)}
\def\ilf{\psi(M)}
\def\msun{M_\odot}
\def\zsun{Z_\odot}
\def\met{[M/H]}
\def\vi{(V-I)}
\def\mtot{M_{\rm tot}}
\def\mhalo{M_{\rm halo}}
\def\pp{\parshape 2 0.0truecm 16.25truecm 2truecm 14.25truecm}
\def\la{\mathrel{\mathpalette\fun <}}
\def\ga{\mathrel{\mathpalette\fun >}}
\def\fun#1#2{\lower3.6pt\vbox{\baselineskip0pt\lineskip.9pt
  \ialign{$\mathsurround=0pt#1\hfil##\hfil$\crcr#2\crcr\sim\crcr}}}
\def\ie{{ i.e., }}
\def\eg{{ e.g., }}
\def\etal{{et al.\ }}
\def\etalc{{et al., }}
\def\kpc{{\rm kpc}}
 \def\Mpc{{\rm Mpc}}
\def\mh{\mass_{\rm H}}
\def\mmax{\mass_{\rm u}}
\def\ml{\mass_{\rm l}}
\def\bc{f_{\rm cmpct}}
\def\br{f_{\rm rd}}
\def\kmsec{{\rm km/sec}}
\def\ibl{{\cal I}(b,l)}
\def\dmax{d_{\rm max}}
\def\dmin{d_{\rm min}}
\def\mbol{M_{\rm bol}}
\def\kms{{\rm km}\,{\rm s}^{-1}}

\lefthead{Graff et al.}
\righthead{Velocity structure of LMC}
\title{The velocity structure of LMC Carbon stars: young disk, old disk, and perhaps a seperate population}
\author{David S. Graff and Andrew P. Gould}
\affil{Departments of Astronomy and Physics, The Ohio State University, Columbus, OH 43210, USA}
\authoremail{graff.25@osu.edu, gould@payne.mps.ohio-state.edu}
\author{Nicholas B. Suntzeff and Robert A. Schommer}
\affil{Cerro Tololo Inter-American Observatory, Casilla 603, La Serena, Chile}
\authoremail{nsuntzeff, bschommer@noao.edu}
\and
\author{Eduardo Hardy}
\affil{National Radio Astronomy Observatory, Casilla 36-D, Santiago, Chile}
\authoremail{ehardy@nrao.edu}
\submitted{Submitted to ApJ 20 Oct. 1999; Accepted 24 Feb. 2000}

\begin{abstract}
We analyze the velocity residuals of 551 carbon stars relative to a
rotating-disk model of the inner $\sim 70\,\rm deg^2$ of the Large
Magellanic Cloud (LMC).  We find that the great majority of the stars
in this sample are best fit as being due to two different populations,
a young disk population containing 20\% of the stars with a velocity
dispersion of $8\,\kms$, and an old disk containing the remaining
stars with a velocity dispersion of $22\,\kms$.  The young disk
population has a metallicity $\sim 0.25$ dex higher than the old disk.

With less certainty, the data also suggest at the $2\sigma$ level that there may
be a third kinematically distinct population that is moving towards us
at 30 km/sec relative to the LMC, consistent with measurements of 21
cm velocities.  If real, this population contains about 7\% of the
carbon stars in the sample.  It could be a feature in the disk of the
LMC or it could be tidal debris in the foreground or background.  If
it is tidal debris, this population could account for some or all of
the microlensing events observed towards the LMC.
\end{abstract}


\section{Introduction}
\setcounter{footnote}{0}
\renewcommand{\thefootnote}{\arabic{footnote}}

Carbon stars are an important tracer of the kinematics of the disk of 
the Large Magellanic Cloud (LMC).  Kunkel \etal
(\cite{kunkel}) have analyzed the velocities of carbon stars in the outer 
LMC disk. Hardy, Schommer, \& Suntzeff (\cite{ssh}) have measured the radial
velocities of 551 carbon stars in the inner $\sim 70\,\rm deg^2$ of the LMC
and fit these velocities to a disk model.  Here we focus on the
residuals to this disk solution in order to isolate different kinematic
components of the LMC carbon star population.

The Milky Way disk has a multi-components structure that may be describable
as a kinematically cold thin disk and a hotter thick disk as originally
advocated by Gilmore \& Reid (1983), or may comprise a continuum of
structures of increasing thickness (Norris \& Ryan 1991).
The LMC provides a unique laboratory to study the kinematic substructure
of a quite different galaxy that also has disk kinematics.  By studying 
this substructure, we can eventually learn about the relations between
formation history, disk heating, and enrichment in a non-Milky Way setting.

Schommer \etal (1992) and Hughes, Wood, \& Reid (1991) have presented
data suggested that older components of the
LMC disk are kinematically hotter than younger components, although these
studies are limited by poor statistics.  In this paper, we analyze
a much larger radial-velocity sample to extract a substantially more detailed 
picture  of the  LMC's disk substructure.  From both conceptual and practical
standpoints, the analysis of the Carbon star radial velocities is best
divided into two steps.  In the first step  Hardy et al.\ (\cite{ssh}) fit
for the global properties of the disk including its projected rotation
curve and its transverse velocity.  Here we apply the second step and
examine the residuals to that fits in order to extract information about
the kinematic structure of the disk.  This study yields unambiguous
evidence that the LMC disk, like the Milky Way disk, has a multi-component
structure.  We go on to show that, just as with the Milky Way, the colder
disk component is more metal rich than the hotter one.

In addition to determining the structure of the LMC disk, we also search for a non-disk component.  One of the motivations for this research is the microlensing
conundrum.  At present $\sim 20$ microlensing events towards the
Magellanic clouds have been analyzed (Alcock \etal
\cite{macho6yr}; Lasserre \etal \cite{eros2lmc}).  If these microlensing events
are due to halo objects, or {\sc Macho}s, then the detected {\sc
Macho}s make up $10-30$\% of the mass of the halo.  All obvious
astrophysical candidates for halo microlensing have severe problems
(e.g. Graff, Freese, Walker \& Pinsonneault \cite{gfwp})
An alternative hypothesis is that the microlensing events are 
due to lenses within
the LMC (Wu \cite{wu}, Sahu \cite{sahu}).  However, if these lenses
are virialized, they must have a large velocity dispersion (Gould
\cite{gouldvir}).  In that case, we should see this population in the
carbon star velocities, unless the carbon stars do not trace the lens
population (Aubourg \etal 1999).

Another possibility is that the observed microlensing is due to an
unvirialized foreground or background population of lenses, such as
a tidal streamer (Zhao \cite{zhao}; Zaritsky \& Lin (1997);
Zaritsky \etal \cite{zsthl99}).
In this case, we would expect the velocities of the lenses to be
different from those of LMC stars.  Again, we should see this population in
the carbon star velocities, unless the carbon stars do not trace the
lens population or unless, by coincidence, the lens population has the
same radial velocity as the main LMC population.  We find that the data provide evidence at the $2\,\sigma$ level for additional
velocity structure that could be due to an unvirialized foreground or
background population.  While this detection cannot be regarded as compelling,
the problem of explaining the observed microlensing events by other routes
has proven so difficult that this proposed solution should be given 
serious consideration: our marginal detection should be checked by a much
larger radial-velocity study.

\section {The Data}

Hardy et al.\ (\cite{ssh}) obtained radial velocities $v$
for 551 carbon stars in 35 fields, each about $0.25\,\rm deg^2$
scattered more or less uniformly over the inner $70\,\rm deg^2$ of the
LMC.  The measurement errors are typically $\sim 1\,\kms$.  Hardy
et al.\ (\cite{ssh}) fit these velocities to a planar, inclined disk
with a circular velocity that is allowed to vary in 5 bins.  Table 1 shows
a summary of the parameters for the solution used in this paper; see Hardy
\etal (\cite{ssh}) and Schommer \etal (\cite{sosh}) for details and
descriptions of the rotation curve parameters and other possible fits.  The
fit adopted here is basically a solid body rotation model (constant dV/dr) out
to 3.5 degrees, a flat rotation curve beyond that (3.5-5.5$^{\circ}$),
a slightly twisting line of nodes ($\Theta$ in Table 1),  an overall
dispersion around the fit ($\sigma$) which is characteristic of an
intermediate to old disk population, and an orbital transverse motion
consistent with the proper motion measures of the LMC (e.g., Kroupa \& Bastian
 1997). The
solution simultaneously fits for the transverse velocity of the LMC
${\bf v}_\perp$ since this gives rise to a gradient in radial
velocities across the face of the LMC with respect to angular
position, $\nabla v = {\bf v}_\perp$.  In this paper we primarily use
the residuals to this fit, $\Delta v$, 
(\S\ 3 and \S\ 4.1) 
but also make use of the
heliocentric radial velocities, $v$ 
 (\S\ 4.2).

\begin{tabular}{ccc} \hline
\multicolumn{3}{c}{Table 1. Rotation Curve Parameters}\\ \hline
V$_{sys}$ & dV/dr& V$_{circ}$ \\

 50 km/s &   21.5 km/s/kpc & 75 km/s \\ \hline
 $<\Theta(PA)>$ & $\sigma$ & V$_{tr}$ \\
 --20$^{\circ}$ & 18-22 km/s & 250 km/sec \\ \hline
\end{tabular}

\section {Detection of two populations}
\label{twopopsec}

A histogram of the residuals $\Delta v$
is shown in Figure \ref{residuals}.  We attempt 
to represent these residuals as various sums of Gaussians of the form
\begin{equation}
\label{twopop}
P(\Delta v) = \sum_{i=1}^n{N_i\over \sqrt{2\pi}\sigma_i}
 \exp\biggl[-{(\Delta v - \overline{\Delta v}_i)^2\over 2\sigma_i^2}\biggr],
\end{equation}
subject to the constraint $\sum_i N_i = 551$.  Here $n$ is the number
of Gaussian components, and for each component $i$, $N_i$ is the
number of stars, 
 $\overline{\Delta v}_i$ 
is the mean residual velocity, and
$\sigma_i$ is the dispersion.  We fit the velocity residuals to these
functional forms by adjusting the parameters to maximize the log
likelihood estimator,
\begin{equation}
\label{likelihood}
\ln L= \sum_{k=1}^{551}\ln[P(\Delta v_k)] \, .
\end{equation}
This is equivalent to a $\chi^2$ minimization measurement in the Poisson limit 
of infinitely small bin size.  Probabilities can be inferred from the log 
likelihood estimator by comparing likelihoods
to the solution with maximum likelihood and using the relation
\begin{equation}
\label{chi2}
\Delta \chi^2 = -2 \Delta \ln L \, .
\end{equation}

	Figure \ref{residuals} shows fits to the (unbinned) residuals
using a single Gaussian (with two free parameters) and a double
Gaussian.  In the latter fit, we impose the physically plausible
additional constraint 
 $\overline{\Delta v}_1=\overline{\Delta v}_2$, 
so there
are a total of 4 free parameters.  The double Gaussian solution has
20\% of the stars in a thin disk population with a velocity dispersion
of $8\,\kms$, and the remaining 80\% of the stars in a thicker disk
population with a velocity dispersion of $22\,\kms$.  The improvement
is $\Delta\chi^2=20$ for the addition of two degrees of freedom, i.e. a
statistical significance of $1-\exp(-\Delta\chi^2/2)\sim 1 -
10^{-4.3}$.  Thus, LMC carbon stars are better represented as two
populations than one.  However, this does not prove that we have
detected two distinct populations.  It could also be that there are a
continuum of populations with a range of dispersions from below 8 to
above $22\,\kms$.  Nevertheless, for clarity of discussion, we will
refer to two discrete populations.

\subsection {Metallicity of the two populations}

Costa \& Frogel (\cite{frogel}) (CF) published $RI$ photometry of 888
LMC carbon stars and 204 with infrared $(JHK)$ photometry.  Within
this sample, 103 of the stars 
 that
have infrared photometry had velocities measured by Hardy \etal (\cite{ssh}).  CF
showed that the infrared colors differ between samples of carbon stars
from the Milky Way, the LMC, and the SMC.  The carbon stars in the
three galaxies can be fit by
\begin{equation}
\label{metallicity}
(J-H)_0=0.62(H-K)_0+\zeta
\end{equation}
with $\zeta \approx \{0.72,0.67,0.60 \}$ respectively for the Galaxy,
the LMC, and the SMC.  Cohen \etal (\cite{cfpe}) suggested that this shift in
colors is due to a metallicity related blanketing effect,  
in which case $\zeta$ can be used as a metallicity indicator.  
As can be seen in 
 Figure 5 
of CF, there is substantial scatter in the 
color-color relations compared to the differences among the three galaxies.
Thus, this metallicity indicator cannot reliably determine the
metallicity of an individual carbon star: it should be used only as a
statistical estimator for stellar populations.

Even though the metallicities of carbon stars in the three galaxies
are unknown, if we assume that ${\rm [Fe/H]}\sim\{ 0,-0.4, -0.8 \}$ for the three galaxies, 
we can make a rough calibration of this metallicity indicator:
\begin{equation}
\label{zetacalib}
\delta {\rm [Fe/H]} \approx 6.7 \, \delta \zeta \, .
\end{equation}
This relation should be taken only as 
rough estimate.  However, one can be more confident of the
relative {\it order} of the metallicities of carbon stars in the three
galaxies, and hence $\zeta$ can robustly distinguish between a
high-metallicity population and a low-metallicity population.

We find that the metallicity indicator $\zeta$ is different for high
velocity-residual stars than for low velocity stars.  Specifically,
for stars with $|\Delta v| < 10 $ km/sec, we find $\zeta=0.678 \pm
0.007$ while for $|\Delta v| > 10$ km/sec, we have $\zeta=0.662 \pm
0.005$.  These two values of $\zeta$ are different at the 93\%
confidence level.  However, since most of the ``low velocity''
stars chosen this way are actually from the more numerous thick-disk
velocity sample, dividing up the sample in this way is not the best way to
measure the metallicity difference.  
To isolate the thin and thick disks, we modify
equation (\ref{twopop}) to read 
\begin{equation}
\label{twopopp}
P(\Delta v) = \sum_{i=1}^n{N_i\over \sqrt{2\pi}\sigma_i}
 \exp\biggl[-{(\Delta v - \overline{\Delta v_i})^2\over 2\sigma_i^2}\biggr]
 \exp\biggl[-{(\zeta - \bar \zeta_i)^2\over 2\sigma_\zeta^2}\biggr],
\end{equation}
where $\bar\zeta_i$ is the mean value of $\zeta$ for each population and
$\sigma_\zeta=0.044$ is the observed dispersion of $\zeta$ in the sample for 
the 103 stars with velocities and infrared data.  Note that for stars 
without infrared data, the last term is simply set to unity.  We then find
$\bar\zeta_1= 0.663\pm 0.04$, $\bar \zeta_2=0.700\pm 0.16$, and
$\bar\zeta_2-\bar\zeta_1=0.037\pm 0.017$, i.e. a $2\,\sigma$ difference,
which corresponds to $\Delta {\rm [Fe/H]}\sim 0.25$.

Given the combination of different velocities and different
metallicities, we claim that we have detected either two different disks
within the LMC representing different ages of stellar populations or
a continuous distribution of disk populations with a range of ages.  In 
either case, the younger populations have higher metallicity and lower velocity
dispersion.

\subsection{No virialized lenses}

Gould (\cite{gouldvir}) showed that for microlensing within a
virialized disk, the microlensing optical depth is
\begin{equation}
\label{vir}
\tau=2 \frac{\langle v^2 \rangle}{c^2} \sec^2 i
\end{equation}
where $i$ is the angle of inclination of the disk with respect to the
line of sight, $30-40 ^\circ$ in the case of the LMC.  In the case of
the carbon stars, the total velocity dispersion is $21\,\kms$
and thus the optical depth due a virialised stellar population
traced by the carbon stars is $\la 2\times 10^{-8}$, much smaller than that measured by the MACHO experiment
(Alcock \etal \cite{macho6yr}) of $1.2 ^{+0.4}_ {-0.3} 
\times 10^{-7}$.  Thus, the 
virialized population traced by carbon stars cannot account for microlensing.  
However, a virialized population too old to be traced by carbon stars would 
not be seen in our data (Aubourg et al.\ 1999). 

\subsection{Conclusion}

We have explicitly assumed that hotter, more metal poor population is older than the younger, metal rich population in analogy with the Milky Way, even though the LMC may have a different disk heating mechanism than the Milky Way.  The age-velocity dispersion relation has been confirmed previously by Hugues, Wood \& Reid (\cite{hughes}) and Schommer \etal (\cite{sosh}).  Since we detect a metallicity difference based on our infrared colors within this population, we also determine that some noticeable metal enrichment occured during the Carbon star formation epoch.

The velocity dispersion of the thick disk component, $22\, \kms$, is much higher than the thin disk, and is close to the velocity dispersion of the oldest objects measured in the LMC, $\sim 30 \kms$ (Hughes, Wood \& Reid 1991, Schommer \etal 1992).  Thus, we can show that the bulk of disk heating occurred during the Carbon star formation epoch.

\section {Search for a Kinematically Distinct Population}

The analysis of Gould (\cite{gouldvir}) only applies to virialized
populations.  It is still possible that an unvirialized population of
stars could be causing microlensing.  Such a population might be a
streamer of stellar material pulled out by tidal interactions between
the LMC and the Milky Way, or between the LMC and the SMC (Zhao \cite{zhao}).
Zaritsky \& Lin (\cite{zl97}) claimed that they may have seen such a
streamer in LMC clump giants.  This paper caused numerous
counter-arguments which are summarized and debated in (Zaritsky \etal
\cite{zsthl99}).  

Ibata, Lewis \& Beaulieu (\cite{ilb}) examined the velocities of
40 clump giants in the LMC of which 24 were candidate
foreground stars according to the criteria of Zaritsky \& Lin
({\cite{zl97}}).  Ibata et al.\ (1998) 
found no difference in the mean velocities of
the candidate foreground stars and the other clump stars and concluded
that these stars did not form a separate kinematic population from the
LMC.  Zaritsky \etal (\cite{zsthl99}) confirmed the
results of Ibata et al.\ (1998) 
using a much larger sample of 190 candidate foreground
clump stars.  However, the carbon-star sample that we analyze here is 
potentially more sensitive to the presence of tidal streamers than
either of these two clump-star samples, in part because it is larger
(551 stars) and in part because the velocity errors are much smaller
($\sim 1\,\kms$).

\subsection{Search for third population in disk-fit residuals}

We search the data for a non-virialized, kinematically distinct
population (KDP) in two different ways.  First, we fit the residuals
to the disk solution to the sum of three Gaussians, two representing
the LMC, and one for the KDP.  That is, we apply equation
(\ref{twopopp}) with $n=3$.  We find a solution which is somewhat
better than the two Gaussian fit, $\Delta\chi^2=8$ for a change of 4
degrees of freedom.  The off-center KDP peak is found to be moving
towards us at $27\,\kms$ relative to the bulk of the LMC and to
contain 63 stars, about 10\% of the total.  Thus, the data suggest
that there may be a KDP, but at a statistically weak level of
confidence.  A Monte Carlo simulation was performed to verify the
statistical confidence (details of which are described in \S\ 4.3)
which showed that this third bump is only present at the 75\%
confidence level.  The fit to the third bump is shown in Fig
\ref{threepopfig}.

\subsection{Search for a third population in velocities}

	In the model considered in the previous section, the KDP stars
have a common motion {\it relative to the LMC}.
Possibly, the KDP stars are moving steadily away from the LMC disk, or
are not associated with the LMC disk.  The KDP should be seen in
the original heliocentric radial velocities $v$ better than it is seen
in the disk-fit residuals $\Delta v$.  We therefore fit the data to a
functions of the form
\pagebreak
\begin{eqnarray}
\label{threepopv}
P(v,\Delta v)=& \sum_{i=1}^2{N_i\over \sqrt{2\pi}\sigma_i}
 \exp\biggl[-{(\Delta v - \overline{\Delta v_i})^2\over 2\sigma_i^2}\biggr]
\exp\biggl[-{(\zeta - \bar \zeta_i)^2\over 2\sigma_\zeta^2}\biggr]
\nonumber \\ & 
+ \,\,
{N_{\rm KDP}\over \sqrt{2\pi}\sigma_{\rm KDP}}
 \exp\biggl[-{(v - (\overline{v}_{\rm KDP}+A_{x}\theta_x + A_{y}\theta_y)^2)
\over 2\sigma_{\rm KDP}^2}\biggr] \nonumber \\ & 
\,\exp\biggl[-{(\zeta - \bar \zeta_{\rm KDP})^2\over 2
\sigma_\zeta^2}\biggr],
\end{eqnarray}
where $(\theta_x,\theta_y)$
is its angular position on the sky, and $A_{x}$ and $A_{y}$ are
planar coefficients for the heliocentric velocity distribution of the
KDP.  This equation is similar to 
 equation 
(\ref{twopopp}) but we have replaced
the $\Delta v$ in the KDP terms by $v$, i.e., we fit to the heliocentric
rather than the residual velocities.  The origen of our x-y coordinate
system is at $\alpha=5^h21^m, \delta= -69^\circ17^\prime$, with X
increasing to the east and Y to the north.

	Initially, we set $A_{x}=A_{y}=0$, so that there
are same number of degrees of freedom as in the three-Gaussian fit to the
residuals.  We find no
solution here that has a lower $\chi^2$ than the two-Gaussian
solution, implying that there is no evidence for the
existence of a third population having a common heliocentric
velocity outside the LMC disk.

	We therefore repeat the search, but allow
$A_{x}$ and $A_{y}$ to vary as free parameters.  We find that the likelihood
is then maximized at very low values of the velocity dispersion
$\sigma_{\rm KDP}\la 1\,\kms$.  We reject these soultions as
unphysical, and note that our fitting routines may have been falsely
attracted to them as results of inevitable Poisson noise.

	We then find a solution with 39 stars in the KDP with $\bar v_{KDP}
= 16.4\,\kms$, $A_x = 2.6\,\kms\,\rm deg^{-1}$, $A_y = 4.9\,\kms\,\rm
deg^{-1}$, $\sigma_{\rm KDP}=5\,\kms$, and $\zeta_{\rm KDP}=0.673$.
Relative to the two-Gaussian solution, this KDP solution has
$\Delta\chi^2=16$ for 6 additional parameters.

Figure \ref{kdpfig} shows the residuals of the LMC stars with respect
to the KDP.  The KDP is shown as the strong peak of points around
residual 0.  Other small peaks are due to the clumped distribution of
our stars in angle, and are not significant.

There are not enough stars in the KDP to significantly determine if
the KDP covers the entire face of the LMC or has a patchy distribution.

\subsection{Monte Carlo}

	While the probability that any randomly chosen plane will
come within $\sim 5\,\kms$ 
 of a significant fraction of our sample stars
is small (and well represented by the
$\chi^2$ test), there are a large number of independent planes that can
be compared to the data.  To obtain a more accurate assessment
of the statistical significance of this detection, we perform a set of Monte
Carlo simulations.  In each simulation, we draw velocities randomly from
the two-Gaussian distribution of disk residuals found in \S\ 3.  We then
search for a KDP in the resulting heliocentric velocities in the same
way we did for the actual data in \S\ 4.1 and \S\ 4.2.  In order to make the
simulations tractable, we ignore metallicity information.  This simplification
is justified by the fact that the metallicity of the KDP measured in 
\S\ 4.2 is not significantly different from the ``young disk'' component.
If metallicity is ignored then the external-plane solution shows an
improvement of $\Delta\chi^2=14$ for 5 additional parameters, which is
formally significant at the 98\% level.  However, we find that out of
407 simulations, there is  $\Delta\chi^2\geq 14$ in 26 cases.  Hence, our
detection is significant only at the 94\% level, roughly equivalent to
$2\,\sigma$.

\subsection{ Evidence of the KDP in Other LMC components}

Given the intriguing signal we see in the C star velocities, but also
the marginal level of significance, it is worth exploring other possible
signs of the KDP. One such tracer is the 21cm gas emission, mapped, e.g.,
by Luks \& Rohlfs (\cite{lh}), and Kim \etal (\cite{kim}). Luks \& Rohlfs
note that a lower velocity component (``L-component'') contains
about 19\% of the HI gas in the LMC, is separated from the main velocity
component by $\sim$30 km/s. Although Kim \etal (\cite{kim}) do not specifically
comment on such a component in their paper based on higher spatial resolution
HI imaging,  a similar signal seems evident in their position-velocity
maps (e.g., 
 Figs.\ 
7a and 7b in their paper) at RA 05:37 - 05:47 and DEC -30
to -120 arcmin.  The standard interpretation of this substructure in
gas is that it is due to hydrodynamic effects on gas within the LMC
disk.  However, the correlation of the gas velocity ``L-component''
with the stellar KDP suggests that the gas may be outside the LMC disk.

An intriguing but somewhat more ambiguous signature may be evident in
the CH star velocities of 
 Cowley \& Hartwick 
(\cite{ch}). Velocities
for a sample of $\sim$80 CH stars show a low velocity asymmetric tail,
consistent with a component at $\sim$20 km/sec lower systematic
velocity. 
 Cowley \& Hartwick (1991)
even suggest that one explanation of
this population is that it is a result of an earlier violent tidal
encounter 
 between
the LMC-SMC system and the Milky Way. The small sample
statistics and asymmetric spatial distribution of these stars make a
more detailed exploration difficult.

\section{Microlensing Interpretation}

We may have detected a kinematically distinct population of carbon stars 
in the direction of the LMC.  If real,  this population could be either
a structure within the LMC disk or tidal debris that is well separated from
the disk and hence either in front of or behind the LMC.  If it is well
separated from the LMC, then it would give rise to microlensing: either
it would be in front of the LMC and so would act as lenses, or it would be
behind the LMC and would act as sources.

The microlensing optical depth due to a thin sheet of stellar
matter with density $\Sigma_1$ and the LMC with density $\Sigma_2$
separated by a distance $D$ which is small compared to the distance
from the 
 Sun
to the LMC is:
\begin{equation}
\label{tau}
\tau_{\rm KDP}  = 
\frac{4 \pi G}{c^2} D \frac{\Sigma_1 \Sigma_2}{\Sigma_1+\Sigma_2}.
\end{equation}

The distance between the two sheets, $D$, cannot be determined from
velocity data alone.  However, since the two sheets must have similar
velocities, the tidal tail cannot be a random interloper in the halo,
but must be somehow related to the LMC.  Lacking further information,
we make the somewhat ad hoc assumption that the material in the tidal
tail has been moving away from the LMC at a constant velocity of $30\,
\kms$ since close tidal encounter between the LMC and the SMC, 200 Myr
ago (Gardiner \& Noguchi \cite{gn}).  In that case, we have
\begin{equation}
\label{D}
D \sim    v_{\rm KDP} \times 200\, {\rm Myr} \sim 5\,{\rm kpc}.
\end{equation}
In fact, it is likely that the foreground object has had its velocity
substantially changed by gravitational interaction with the LMC, and
to a lesser extent, the SMC and the Milky Way, so this calculation
only indicates that the object could have moved several kpc from the
LMC in the past 200 Myr.  All the results of this section will hold if the
object is several kpc either in front of or behind the LMC. 

The total surface mass density, $\Sigma_1 + \Sigma_2$, can be
estimated from the observed surface brightness of
the LMC, which is $R\sim 21.2$
mag arcsec$^{-2}$ (De Vaucouleurs \cite{dvc}) near the center.  
If we assume a mass to
light ratio of 3 (in solar units), this corresponds to a total surface
mass density of 300 $\msun$ pc$^{-2}$.

It is possible that the surface densities of the disk and KDP populations are not traced by carbon stars.  Still, lacking further information, we estimate the optical depth by setting 
$\Sigma_1/\Sigma_2=39/(551-39)$ according to the solution of \S\ 4.2,
we obtain
\begin{equation}
\label{result}
\tau= 6\times 10^{-8}\,{D\over 5\,\kpc}.
\end{equation}
This optical depth is substantially larger than the optical depth due
to a virialized disk population traced by the carbon stars ($\la
2\times 10^{-8}$).  It is consistent with the value
observed by the {\sc Macho} collaboration (Alcock \etal 2000).  There
could be more tidal material which we have not found in this search
because its velocity is by chance too close to the velocity of the
LMC, and which would raise the optical depth.  If $D$ were greater
than 5 kpc, then $\tau_{\rm KDP}$ would rise proportionately.

The transverse motion of such a population with respect to the LMC is
probably 70 $\kms$, the circular orbital velocity of the LMC.  To
calculate the typical transverse velocity in a microlensing event,
this velocity should be added in quadrature to all the other sources
of transverse velocity.  The stars in the LMC are orbiting about the
LMC center with a transverse motion of 70 $\kms$ at 4 kpc (Kunkel
\etal \cite{kunkel}; Hardy \etal \cite{ssh}).  The LMC system has a
transverse velocity with respect to the 
 Sun
of some 250 $\kms$ (Hardy
\etal \cite{ssh}) which will translate to a projected transverse
motion of 25 $\kms$ (at 5 kpc from the LMC).  Adding these velocities
in quadrature, the derived typical transverse velocity of a
microlensing event is 100 $\kms$, in which case the typical mass of a
lens is
\begin{equation}
M\sim 0.13\, \msun\, \biggl({D\over 5\,\kpc}\biggr)^{-1}. 
\label{Mest}
\end{equation}
 This is significantly below the
mean mass of stars in the neighborhood of the Sun (e.g. Gould, Bahcall, \& 
Flynn 1997), but the LMC may have a different mass function.  However, it
is important to recognize that if $D$ is made larger so as to account
for more of the optical depth, then the mean mass is driven lower
\begin{equation}
M\sim 0.075\,M_\odot\,\biggl({\tau\over 1\times 10^{-7}}\biggr)^{-1}.
\label{Mesttau}
\end{equation}

\section{Conclusion}

We report two primary new results, one with high statistical confidence, one which is shakier, but perhaps more interesting if true.  We show that Carbon stars in the LMC are divided into a hot and cold population, with a clear difference in metallicity between the two populations.  Thus, we show that the epoch of LMC disk heating had to occur during the Carbon Star formation epoch.

We also show with less confidence the existence of a third population, outside the LMC.  If this population is real, it suggests that some fraction of the Carbon stars in the LMC are not in the disk, and thus could explain microlensing events.  Although at present the statistical significance of this detection is not enviable, this result is still the best extant solution to the microlensing conundrum.  The microlensing conundrum poses such a difficult problem that several extreme explanations have been proposed including mirror matter and cosmological populations of population III white dwarfs (Graff, Freese, Walker \& Pinnsoneault 1999, and references therin).  The kinematically distinct population is unique amongst these explanations in that it in not only {\it allowed} by the data, but even supported by the data at the 95\% confidence level, and requires no modifications to the standard models of Particle Physics or Cosmology.  We thus present it as the strongest explanation of LMC microlensing.

\acknowledgements

Work at Ohio State was supported in part by grant AST 97-27520 from the NSF.

\pagebreak

\section*{Figure Captions}

\begin{figure}[ht]
\caption{The residuals of the carbon stars with respect to the LMC
disk fit.  The grey line is the best fit single gaussian.  The black
line is the best fit two gaussian model of eq. (\ref{twopop}).}
\label{residuals}
\end{figure}

\begin{figure}[ht]
\caption{A fit to the residuals with three gaussians.  Although the
third peak is not significant in the fit to the residual, it is shown
to be statistically significant when searched for in velocities, and
shows the location of the KDP.}
\label{threepopfig}
\end{figure}

\begin{figure}[ht]
\caption{
The residuals of the stellar velocities with respect to the KDP.  The
KDP stands out as a strong peak near residual = 0.}
\label{kdpfig}
\end{figure}

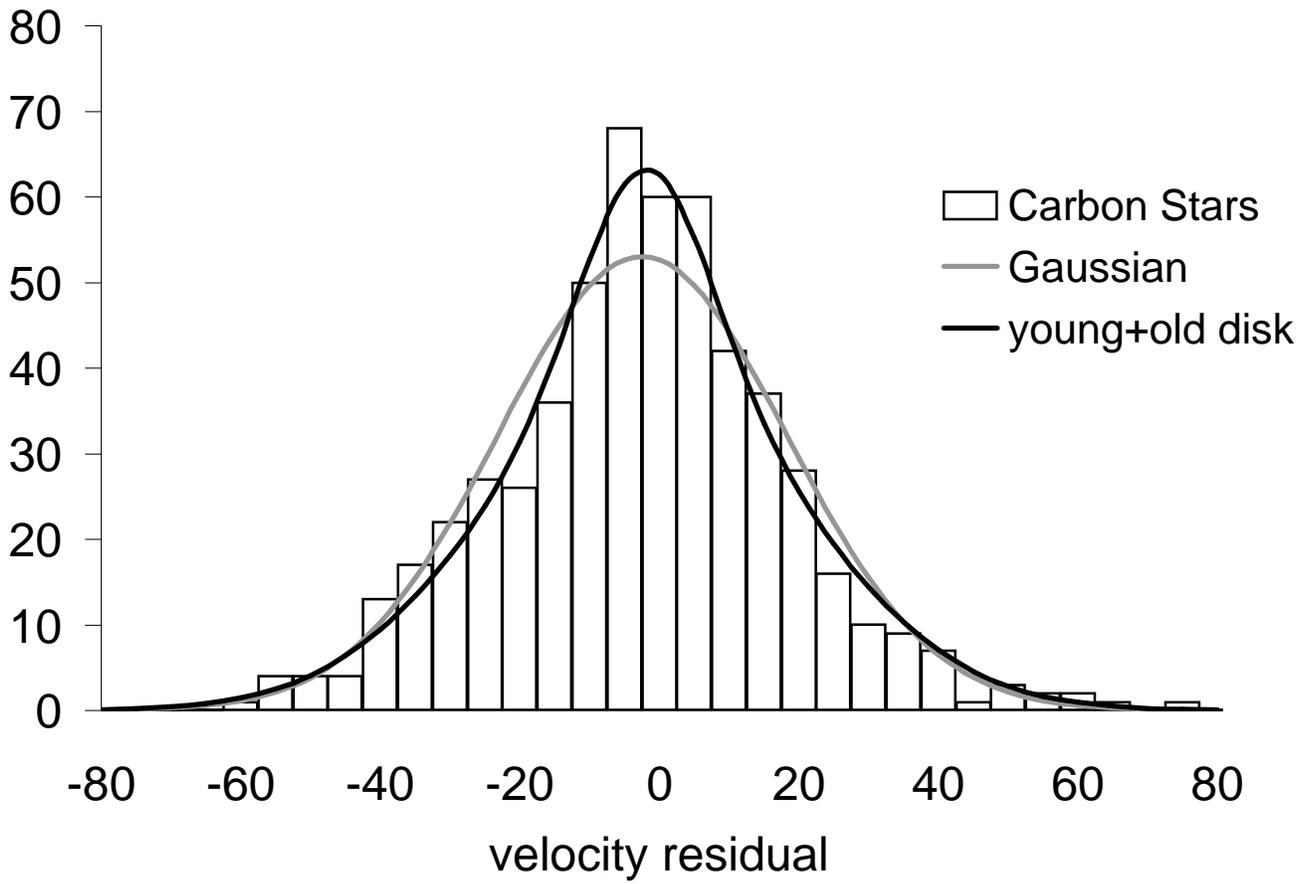
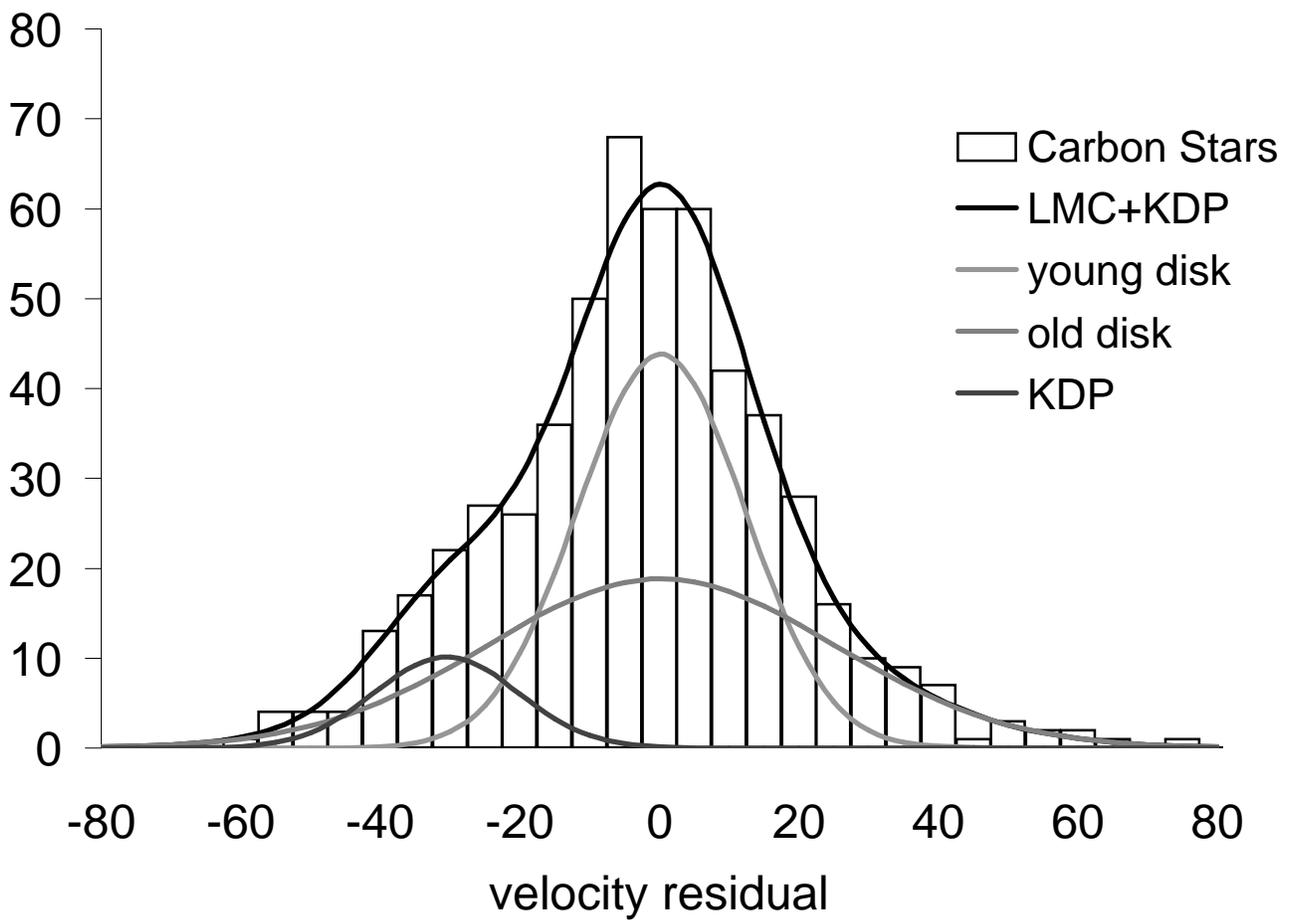
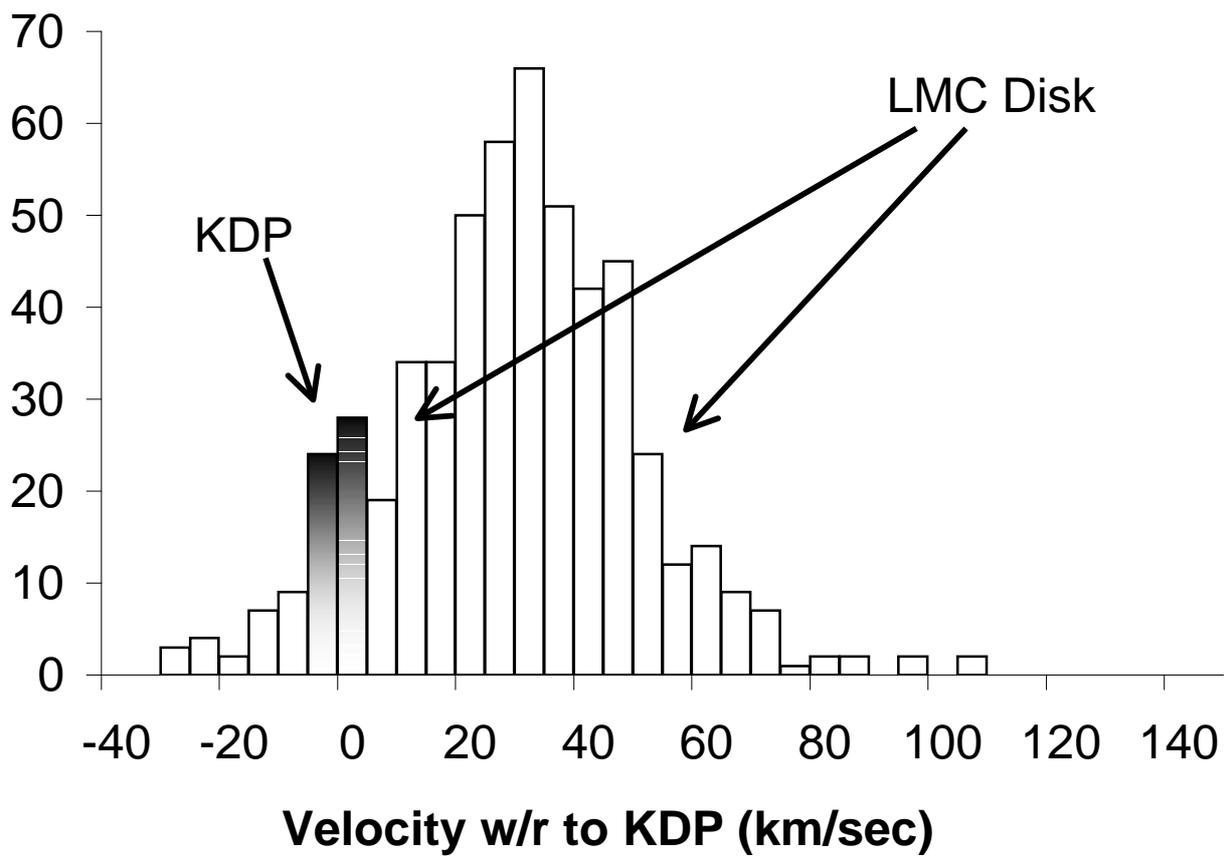
\end{document}